\documentclass[preprint,showpacs,preprintnumbers,amsmath,amssymb,nofootinbib]{revtex4}

\usepackage{etex}
\usepackage{graphicx}
\usepackage{dcolumn}
\usepackage{bm}
\usepackage{amssymb,amsthm,amscd,amsbsy,array}
\usepackage{amsmath,dsfont}
\usepackage{soul} 
\usepackage{graphics,xcolor}

\numberwithin{equation}{section}

\newcommand{\half}{{\frac{1}{2}}}
\def\2{{\frac{1}{2}}}

\def\parag{\hfil\break} 
\def\kikezd{\parag\underbar}
\def\bA{{\bm{A}}}

\def\p{{\partial}}

\def\bR{{\mathds{R}}}

\def\cL{{\mathcal{L}}}
\def\LBB{{\cL_{F}}}
\def\KGL{{L}_{V}}  

\def\bnabla{\mbox{\boldmath$\nabla$}}
\def\br{{\bm{r}}}

\def\bE{{\bm{E}}}
\def\bB{{\bm{B}}}
\def\bC{{\bm{C}}}

\def\bnabla{{\bm{\nabla}}}

\def\bS{{\bm{S}}}

\def\bj{{\bm{j}}}

\def\beq{\begin{equation}}
\def\eeq{\end{equation}}
\def\beqa{\begin{eqnarray}}
\def\eeqa{\end{eqnarray}}

\def\barray{\left(\begin{array}}
\def\earray{\end{array}\right)}
\def\barraynb{\begin{array}}
\def\earraynb{\end{array}}

\def\smallover#1/#2{\hbox{$\textstyle\frac{#1}{#2}$}} %

\def\dAlem{\vcenter {
    \hbox {\vrule height8pt width0.4pt depth0.0pt
           \vrule height8pt width7.2pt depth-7.6pt
           \vrule height8pt width0.4pt depth0.0pt
           \kern-8pt
           \vrule height0.4pt width8pt depth0.0pt
          \,}}}

\usepackage{color}
\usepackage[colorlinks=true, pdfstartview=FitV, linkcolor=blue, citecolor=blue, urlcolor=blue]{hyperref}

\newcommand{\gb}{\colorbox{green}}

\newenvironment{redtext}{\color{red}}{\ignorespacesafterend} 
\newenvironment{bluetext}{\color{blue}}{\ignorespacesafterend}

\newcommand{\bblue}{\begin{bluetext}} 
\newcommand{\eblue}{\end{bluetext}} 
\newcommand{\bred}{\begin{redtext}}
\newcommand{\ered}{\end{redtext}}

\def\?{{\;\gb{\;\large ?\;}\;}}

\begin{document}

\preprint{arXiv:1608.08573v4}

\title{Duality and helicity~: the photon wave function approach
}

\author{
M. Elbistan$^{1}$\footnote{mailto:elbistan@impcas.ac.cn.},
P. A. Horv\'athy$^{1,2}$\footnote{mailto:horvathy@lmpt.univ-tours.fr},
 P.-M. Zhang$^{1}$\footnote{e-mail:zhpm@impcas.ac.cn},
}

\affiliation{
${}^1$ Institute of Modern Physics, Chinese Academy of Sciences, Lanzhou, (China) 
\\
${}^2$ Laboratoire de Math\'ematiques et de Physique
Th\'eorique,
Universit\'e de Tours,  (France)
}

\date{\today}

\begin{abstract}
The photon wave equation proposed in terms of the Riemann-Silberstein vector is derived from a first-order Dirac/Weyl-type action principle. It is symmetric w.r.t. duality transformations, but the associated Noether quantity vanishes.
Replacing the fields by potentials and using instead a quadratic Klein-Gordon-type Lagrangian allows us to recover the  double-Chern-Simons expression of conserved helicity and is shown to be equivalent to recently proposed alternative frameworks. Applied to the potential-modified theory
the Dirac/Weyl-type approach  yields again zero conserved charge, whereas the Klein-Gordon-type approach applied to the original setting yields  Lipkin's ``zilch''. 
\\
\end{abstract}


\maketitle

\tableofcontents

\section{Introduction}

The century-old problem of symmetry of the vacuum Maxwell equations under duality transformations \cite{HeLa},
\beq
\bE\to \cos \theta\, \bE+\sin \theta\, \bB, 
\qquad
\bB\to - \sin \theta\, \bE+\cos \theta\, \bB,
\label{emdual}
\eeq 
has attracted considerable recent attention
\cite{AfSt,BBN,BaCaYa,Camer,FeCo,Manuel,EDHZ-heli,Agullo}.
In particular, the  associated Noether quantity  is the (optical) \emph{helicity} \cite{Calkin}, expressed as the integral of two Chern-Simons terms
\beq
\chi 
=\frac{1}{2}\int_{\bR^3}\!(\bA\cdot \bB - \bC\cdot\bE)\,d^3\br\,,
\label{optihel}
\eeq
for the electromagnetic potential $\bA$ and its dual $\bC$ \cite{AfSt,DeTe}.

While it is possible to obtain (\ref{optihel}) within standard Maxwell theory using Noether's theorem, such a derivation is somewhat complicated, since the Maxwell Lagrangian is not invariant under (\ref{emdual}) \cite{Calkin, DeTe}. Dual-symmetric Lagrangians, have been proposed in \cite{BBN, Camer}. 
The aim of our ``Variation on the Duality/Helicity Theme'' here is to shed some new light on this old subject. We use suitably modified versions of the photon wave function --- a concept which has, admittedly, a long history. The Dirac-type transcription of Maxwell's electromagnetism has been considered by E. Majorana as early as in 1928 \cite{Majorana};
 see \cite{DBN,BaCaYa,Camer} for further details.

We start with  Dirac/Weyl-type first-order equations advocated by Iwo and Zofia Bialynicki-Birula \cite{BiBi}, eqn. (\ref{bFeqns}) below. They are  duality-symmetric however the associated helicity vanishes. Our main result here, presented in sec. \ref{potwf}, is to show that the non-trivial expression, (\ref{optihel}) above,  can be recovered, though, when the \emph{fields are replaced by potentials} and a \emph{Klein-Gordon-type quadratic action} is used.

We note also that while using the Dirac/Weyl-type first-order Lagrangian in the potential-setting would yield again zero charge,  our quadratic Klein-Gordon-type approach applied to the original framework yields, instead, Lipkin's ``zilches'' \cite{Lipkin,Candlin,Kibble}, --- which is thus another manifestation of duality symmetry.

Our results provide us with a nice illustration to  the theorem of Weinberg and Witten \cite{WeWi} on spin and helicity of massless particles. 

\section{The photon wave function}\label{BiBiwf}


 Following Bialynicki-Birula \cite{BiBi} we rewrite the vacuum Maxwell equations as a wave equation reminiscent of Dirac and/or Weyl.
Their starting point is that 
 requiring that the Riemann-Silberstein vector 
\beq
\bm{F}=\frac{1}{\sqrt{2}}\big(\bE+{i}\bB\big)
\label{RSvector}
\eeq 
satisfies the coupled system
\beq
i\,\partial_t\bm{F}=\bnabla\times\bm{F},
\qquad
\bnabla\cdot\bm{F}=0
\label{bFeqns}
\eeq
is  {equivalent} to  the vacuum Maxwell equations with $\varepsilon_0=\mu_0=1$. In terms of the
$3\times3$ rotation matrices in the spin $1$ representation, $\big(S_j\big)_{ab}=-i\,\epsilon_{jab},\, j=1,2,3$,
the first eqn in (\ref{bFeqns}) and its complex conjugate can also be presented  as
\beq
\label{BBeqnR}
\,\partial_t\bm{F}=-\,(\bm{S}\cdot\bnabla)\bm{F}
\qquad\hbox{and}\qquad
\,\partial_t\bm{F}^*=\,(\bm{S}\cdot\bnabla)\bm{F}^*.
\eeq
These two equations are plainly equivalent; note here  the opposite signs. Then, imitating the Dirac procedure (understood intuitively as ``taking the square root
of the Klein-Gordon equation''  \cite{Dirac,BjDr}~), they note that the spin-$1$ rotation matrices satisfy
\beq
(\bm{S}\cdot\bnabla)(\bm{S}\cdot\bnabla)=\bnabla^2
\label{sqrtdelta}
\eeq 
provided that the divergence condition $\bnabla\cdot\bm{F}=0$  holds also. Iterating (\ref{BBeqnR}) shows 
 that each component of the electromagnetic field satisfies the wave equation,
\beq
\big[\p_t^2-\bnabla^2\big]F_i=0.
\label{waveq}
\eeq 
Conversely, (\ref{sqrtdelta}) allows  us to
 take the  ``square root'' of the D'Alembert operator and to posit the two, equivalent equations in (\ref{BBeqnR}), supplemented by $\bnabla\cdot\bm{F}=0$.


The next step is to introduce 
a $6$-component vector and  the $6\times6$ matrices
\beq
\label{BBF+F-}
\mathcal{F}=\barray{c}
\bm{F}_+
\\
\bm{F}_-
\earray\,
\qquad
\rho_1=\barray{cc}
0&1_3
\\
1_3&0
\earray
\qquad
\rho_3=\barray{cc}
1_3&0
\\
0&-1_3
\earray
\qquad
\Sigma^\mu=
\begin{pmatrix}
0 & \overline{S}^\mu\\
S^\mu & 0
\end{pmatrix},
\eeq
$\mu= 0,\dots,3$,
where $
 S^\mu=(1, \bm{S})$ and 
$\overline{S}^\mu=(1, -\bm{S})
$
and to note that putting
\beq
\bm{F}_-=\bm{F}_+^*\,,
\qquad
\bm{F}_{+} = \bm{F} = \frac {1}{\sqrt {2}} (\bE + i\bB)
\label{V-V*}
\eeq
unifies the two eqns (\ref{BBeqnR})  into a 6-component first-order \emph{Dirac-type} equation supplemented with the divergence constraint, 
\begin{subequations}
\begin{align}
\label{BBweqn}
\Sigma^\mu\,\partial_\mu\mathcal{F}&
=0
\\
\bnabla\cdot\mathcal{F}&=0.
\label{Vdivcons}
\end{align}
\label{BB6eqns}
\end{subequations}
We stress that the conjugacy condition (\ref{V-V*}) is \emph{necessary for recovering the Maxwell  theory from the extended one here} \footnote{Our strategy is analogous to what is usually done for the Schr\"odinger Lagrangian, where $\psi$ and $\psi^*$ are first viewed as independent; then one identifies the latter with the complex conjugate of the former after deriving the variational equation. A rigorous mathematical treatment would require using a Lagrange multiplier.}.

The matrix $\rho_3$ acts diagonally but changes the sign of the lower component, allowing us to identify  left and right helicity states as eigenvectors
of $\rho_3$ with eigenvalues $\pm1$.

 The two massless 3-component equations with fixed helicities satisfied  by $\bm{F}_{\pm}$ are uncoupled; they are the spin-$1$ counterparts of the \emph{Weyl equations}, which describe neutrinos and antineutrinos with  spin $1/2$.
$\rho_3$ is the analog of the chirality operator $\gamma^5$;
$\rho_1$ intertwines the helicity components,  $\rho_1\mathcal{F}_{\mp}=\mathcal{F}_{\pm}$.

Now, inspired by the analogy with the Dirac/Weyl system, we propose an action principle for the Dirac-type equation  (\ref{BBweqn}), 
\beq
\label{BBlag}
\LBB=\overline{\mathcal{F}}\,(\Sigma^{\mu}\partial_\mu)\,\mathcal{F}=
\Big(\bm{F}_-^\dagger\,\overline{S}^\mu\partial_\mu \bm{F}_-
\;+\;
\bm{F}^\dagger_+\,S^\mu\partial_\mu \bm{F}_+\Big)\,, 
\qquad
 \overline{{F}}=\mathcal{F}^\dagger \Sigma^0.
\eeq
Our Lagrangian is reminiscent of but still different from the one proposed by Drummond \cite{Drummond2}. 
Treating $\mathcal{F}$ and $ \overline{\mathcal{F}}$ as independent fields, the Euler-Lagrange equations reproduce eqn (\ref{BBweqn}) and its conjugate
 when $ \overline{\mathcal{F}}=\mathcal{F}^\dagger \Sigma^0$ is used.
Expressing in electric and magnetic terms, \vskip-5mm
\begin{subequations}
\begin{align}
\label{BBmax}
{\LBB}
&=
\bE\cdot\big(\partial_t\bE
-\bnabla\times\bB\big)
+\bB\cdot\big(\partial_t\bB+\bnabla\times\bE\big)
\\
\label{BBcont}
&=
\partial_t(\half(\bE^2+\bB^2))+\bnabla\cdot(\bE\times\bB), \end{align}
\label{EBBBLag}
\end{subequations}
shows that ${\LBB}$ is {different} from the usual e.m. Lagrange density $\half(\bE^2-\bB^2)$; 
it is indeed the \emph{divergence of the current ${\cal T}^\mu=(T^{00},T^{i0})$ associated with the \emph{usual} electromagnetic energy-momentum tensor}. ${\LBB}=\p_{\mu}{\cal T}^\mu$ vanishes therefore when the Maxwell equations, (\ref{bFeqns}), are satisfied. 
 
The theory given by (\ref{EBBBLag}) is duality invariant~: the transformation (\ref{emdual}), written  as 
\beq
\mathcal{F} \to e^{-i\theta\rho_3}\mathcal{F},
\label{Vdual}
\eeq
plainly leaves the Lagrange density (\ref{BBlag}) invariant 
 because $\rho_3$ and $\Sigma_\mu$ anticommute, $\{\rho_3, \Sigma^\mu\}=0$, in  analogy with what happens for Dirac/Weyl for spin $1/2$. Then the Noether theorem  provides us with the conserved current,
\begin{eqnarray}
\label{BBNc}
{k}^\mu=
\overline{\mathcal{F}}\,\Sigma^\mu\rho_3\, \mathcal{F}
=\bm{F}_+^\dagger S^\mu \bm{F}_+-\bm{F}_-^\dagger \overline{S}^\mu \bm{F}_-\,,
\qquad
\p_{\mu}{k}^\mu=0,
\end{eqnarray}
which is reminiscent of the chiral current of a massless particle with spin $1/2$.
 However, this current is \emph{identically zero} when the conjugacy condition (\ref{V-V*}) is used,
\beq
k^\mu\equiv0
\quad\Rightarrow\quad
\chi_F=\int\!d^3\br\, k^0=
\int\! d^3\br
\big(\bm{F}_+^\dagger\bm{F}_+-\bm{F}_-^\dagger\bm{F}_-\big)=0.
\label{falsehel}
\eeq
We conclude that the theory  in (\ref{BB6eqns}) is  \emph{unsuitable} to derive helicity, (\ref{optihel}).

We also mention that this theory has further unusual aspects: for example, unlike the wave function in quantum mechanics,  $\mathcal{F}$ in (\ref{BBF+F-}) has \emph{no} gauge degrees of freedom~: the strict gauge invariance of the fields implies strict invariance for $\mathcal{F}$.
Note also that neither the action (\ref{BBlag}) or (\ref{EBBBLag}) nor do the further conserved quantities have correct physical dimension.

\section{A wave function composed of potentials}\label{potwf}


Now we put forward our theory, -- the main result of this paper. Our clue for obtaining  nontrivial dual-symmetry is the observation that the  wave equation (\ref{waveq}) is satisfied also by the  electromagnetic \emph{potentials} when the Lorentz gauge is chosen. We define therefore the new Riemann-Silberstein-type vectors $\bm{V}_\pm$ by \emph{replacing  fields by  potentials in the definitions}  (\ref{RSvector}) and (\ref{bFeqns}), 
\vskip-6mm
\begin{subequations}
\begin{align}
\label{F+-}
\bm{V}_\pm &= \frac{1}{\sqrt{2}}(\bm{A}\,{\pm}\,i{\,}\bm{C}), 
\\[2pt]
\label{potchoices}
\bnabla\times\bm{A}&=-\partial_t \bm{C}\;(=\bB),
\qquad
\bnabla\times\bm{C}=\partial_t \bm{A}\; (=-\bE),
\qquad
A^0=C^0=0.
\end{align}
\label{ptF+F_}
\end{subequations}
Then the conjugacy condition (\ref{V-V*}) --- but now for the potentials~:
\beq 
\bm{V}_-=\bm{V}^*_+\,,
\label{F-F*}
\eeq    
is, once again, built into the theory.
We have also incorporated a double gauge freedom,
$ 
\bm{V}_{\pm}\to \bm{V}_{\pm} +\bnabla{f}\pm {i}\bnabla{g},
$ 
which is \emph{not} that of a usual wave function but  nevertheless legitimates the choice (\ref{potchoices}), which imply 
$ 
\bnabla\cdot\bm{A}=\bnabla\cdot\bm{C}=0.
$ 
I.e., we choose the transverse Coulomb gauge,
cf. \cite{AfSt,BBN,Camer}. Then both of our potentials verify the Lorenz gauge condition 
$ 
\partial_\mu A^\mu=\partial_0 A^0+\bnabla\cdot\bm{A}=0.
$ 
When the Maxwell equations hold,  each component of $\bm{V}_\pm$ satisfies, once again, the free wave equation, (\ref{waveq}), allowing us to \emph{postulate, conversely, new field equations for the new wave functions $\bm{V}_{\pm}$,}  i.e.,
to require that
\begin{subequations}
\label{poteqn}%
\begin{align}
\label{potF}
\,\partial_t\bm{V}_\pm&={\mp}\,\,(\bm{S}\cdot\bnabla)\bm{V}_\pm,
\\
\bnabla\cdot\bm{V}_{\pm}&=0
\label{divcond}
\end{align}
\end{subequations}
hold.

\goodbreak%
Eqns (\ref{poteqn}) are of the first order in the potentials.
Then taking divergences and curls allows us to deduce that  they imply  the vacuum Maxwell equations. 
From  (\ref{poteqn}) we infer also that
$  
\partial_t^2 \bm{V}_\pm-(\bm{S}\cdot\bnabla)(\bm{S}\cdot\bnabla)\bm{V}_\pm
=0
$ which implies, using (\ref{sqrtdelta}),  two (equivalent)  massless  Klein-Gordon equations (D'Alembert equations),
\beq
\label{KGeqns}
\partial_\mu \partial^\mu \bm{V}_\pm\equiv
\Big[\partial_t^2-\bnabla^2\Big]\bm{V}_\pm=0.
\eeq
(\ref{poteqn}) is therefore a square root of the K-G type wave equations satisfied by $\bm{V}_{\pm}$. 
Remembering Klein-Gordon, we note that  (\ref{KGeqns}) derive, after putting  $\bm{V}_+=\bm{V}$ and $\bm{V}_-=\bm{V}^*$, from the manifestly dual-symmetric 
 Lagrangian\footnote {For the relativistic invariance of electromagnetic theory in Coulomb gauge, we refer to \cite{cohen-tannoudji}.}
\begin{eqnarray}
\label{BBKGlag}
{\KGL}&=&\frac{1}{2}(\partial_\mu \bm{V}_-)\cdot(\partial^\mu \bm{V}_+)\,. 
\end{eqnarray}

Further insight is gained by noting that  (\ref{BBKGlag}) is 
in fact  \emph{equivalent} to the one considered in \cite{BBN,Camer},
\beq
\underbrace{\frac{1}{2}(\partial_\mu \bm{V}_-)\cdot(\partial^\mu \bm{V}_+)}_{our \ L_V}=
\underbrace{-\frac{1}{8}\Big[F_{\mu\nu}F^{\mu\nu}+\star F_{\mu\nu}\star F^{\mu\nu} \Big]}_{Barnett \ et \ al - Bliokh \ et\ al} - \underbrace{\frac{1}{4}\p_i\Big(A_{j}\p_j A_{i}+
C_{j}\p_j C_{i}\Big)
}_{surface\ term}.
\eeq
Direct use of (\ref{ptF+F_}b) in (\ref{BBKGlag}) yields a vanishing Lagrangian \cite{BBN,Camer}. Therefore, in analogy with what is done for the Schr\"odinger Lagrangian (see our footnote \# 1 above) and also for the complex, scalar K-G theory, we first consider $\bm{V}_+$ and $\bm{V}_-$ as independent 
 and derive the equations of motion (\ref{KGeqns}) by treating $\bm{V}_\pm$ separately, before inserting the constraint (\ref{F-F*}). The  definitions (\ref{ptF+F_}b) will be used in eqn. (\ref{BBKGNoetherc}) -- (\ref{ourT0i}).
 
\vskip2mm
Turning now to duality,  it is readily seen that (\ref{emdual}), implemented on the potentials as 
$
\bm{A}\to \bm{A}\cos\theta +\bm{C}\sin\theta,\;
\bm{C}\to \bm{C}\cos\theta -\bm{A}\sin\theta\,,
$
i.e.,
\beq
\label{Fdual}
\bm{V}_\pm\to \bm{V}_\pm e^{\mp i\theta},
\eeq
 leaves (\ref{poteqn}) invariant, establishing the duality symmetry of the  system we propose. In fact, the action $S=\displaystyle\int\! d^4x\, {\KGL}$ is manifestly invariant under (\ref{Fdual}). 
The infinitesimal version of the latter, 
$ 
 \delta\bm{V}\,=-i\theta\bm{V},
\;
 \delta\bm{V}^*=\;i\theta\bm{V}^*,
$
allows us to infer the  Noether current
\beqa
j^\mu=
\frac{1}{2}\Big(\partial^\mu\bm{V}\cdot\delta\bm{V}^*+\partial^\mu\bm{V}^*\cdot\delta\bm{V}\Big)
=
\frac{1}{2}\Big((\partial^\mu \bm{A})\cdot\bm{C}-(\partial^\mu\bm{C})\cdot\bm{A}\Big),
\label{BBKGNoetherc}
\eeqa
whose conservation, $\p_{\mu}j^\mu=0$,
can also be checked directly using (\ref{KGeqns}).
The associated conserved charge is the space integral of the zeroth component,
\beq
\chi=\int d^3\br\  \frac{1}{2}\Big(\partial_t \bm{A}\cdot\bm{C}-\partial_t\bm{C}\cdot\bm{A}\Big)=
\int d^3\br\ \frac{1}{2}\Big(-\bE\cdot\bm{C}+\bB\cdot\bm{A}\Big),
\label{truehelicity}
\eeq
by (\ref{ptF+F_}) ---
where we recognize (\ref{optihel}), the ``double Chern-Simons'' expression  of  helicity \cite{Calkin, DeTe, AfSt, BBN, Camer, EDHZ-heli}. Conversely, the charge (\ref{truehelicity}) [i.e. \ref{emdual}] generates the duality action (\ref{Fdual}).

We note also that the space part of  (\ref{BBKGNoetherc}) is [up to a surface term] the \emph{spin density} \cite{BBN}, 
\beq
\bj=\bS = \half (\bE\times\bA+\bB\times\bC).
\label{jcur}
\eeq

The constraint (\ref{F-F*}) does \emph{not} now imply the vanishing of  the helicity in (\ref{truehelicity})~:  one of the factors has been changed into a field strength, cf. (\ref{ptF+F_}).
The integral (\ref{truehelicity}) i.e. (\ref{optihel}) can indeed be evaluated using Fourier transformation to momentum space \cite{AfSt}, showing that the helicity is  proportional to the \emph{difference of the number of left- and right-handed photons},
$ 
\chi=n_L-n_R.
$
Similar formulae hold for the current (\ref{jcur}),
 \cite{BaCaYa}.
 
The  energy-momentum tensor of our theory, 
$T^{\mu\nu}$, is symmetric. 
Spelled out in terms of fields and potentials it is, \begin{subequations}
\label{ourTmn}%
\begin{align}
\label{ourT00}
T^{00} &=\frac{1}{4}\big(\partial_t \bm{A}\cdot \partial_t \bm{A}+\partial_t \bm{C}\cdot \partial_t \bm{C}+\partial_i \bm{A}\cdot \partial_i \bm{A}+\partial_i \bm{C}\cdot \partial_i \bm{C}\big),
\\
T^{0i}&=\frac{1}{2}(\partial_t \bm{A}\cdot \partial^i \bm{A}+\partial_t \bm{C}\cdot \partial^i \bm{C}), 
\label{ourT0i}
\end{align}
\end{subequations}
Its conservation,
$\partial_\nu T^{\mu\nu}=0$, can be checked also directly.  $T^{00}$ and $T^{0i}$
are, up to  surface terms, the usual expressions of the energy density and of the Poynting vector, respectively.

\section{Two more ``variations''} 

(i) Now we shortly discuss our other approaches.
The $\bm{V}_\pm$ could again be unified into a $6$-component system by putting 
$\mathcal{V}=\barray{c}
\bm{V}_+
\\
\bm{V}_-
\earray
$. 
The two helicity components are interchanged by $\rho_1$. 
The two upper equations in (\ref{poteqn}) are also  unified and are 
supplemented with the divergence constraint, 
\begin{subequations}
\begin{align}
\label{F+F-eqn}
\Sigma^\mu\,\partial_\mu\mathcal{V}&
=0,
\\
\bnabla\cdot\mathcal{V}&=0,
\label{Fdivcons}
\end{align}
\label{F+F-6eqns}
\end{subequations}
as in (\ref{BB6eqns})~:
 once again, we get Dirac / Weyl type analogs.
The Lagrangian (\ref{BBKGlag}) can also be written as in (\ref{BBlag}) but with  $\bm{V}_\pm$ replacing $\bm{F}_\pm$\,, 
\beq
\label{goodF+F-}
\cL_{\mathcal{V}}=\,\overline{\mathcal{V}}(\Sigma^{\mu}\partial_{\mu})\mathcal{V}
=
\left(\bm{V}_{-}^{\dag }\overline{S}%
^{\mu }\partial_{\mu }\bm{V}_{-}+\bm{V}_{+}^{\dag }S^{\mu}\partial_{\mu}\bm{V}_{+}\right) ,\ \ \ \ \overline{\bm{V}}=\bm{V}^{\dag }\Sigma^{0},
\eeq
which is again  4-divergence,
\beq 
\cL_{\mathcal{V}}=\partial_{t}\big(\half(\bA^{2}+\bC^{2})\big)+\bnabla \cdot \left(\bA\times \bC\right)
\eeq 
which vanishes when the field equations are satisfied, cf. (\ref{EBBBLag}).  

The Lagrangian (\ref{goodF+F-}) is invariant w.r.t. duality, (\ref{Fdual}), and yields a  Noether current similar to (\ref{BBNc}), 
\begin{eqnarray}
\label{BBNc6comp}
\ell^\mu=\overline{\mathcal{V}}\,\Sigma^\mu\rho_3\, \mathcal{V} =\bm{V}_+^\dagger S^\mu \bm{V}_+-\bm{V}_-^\dagger \bar{S}^\mu \bm{V}_-\,.
\end{eqnarray}
However the current and the to-be helicity  vanish again due to $\bm{V}_+^*=\bm{V}_-$, 
\beq
\ell^\mu\equiv0
\quad\Rightarrow\quad
\chi_V=\int\! \ell^0 d^3\br=\int\!\big(\bm{V}_+^\dagger \bm{V}_+-\bm{V}_-^\dagger \bm{V}_-\big)d^3\br=0.
\eeq 
We conclude that the Dirac-type approach yields, once again, trivial current and charge.

(ii) What would our Klein-Gordon trick yield for the original setting of Section \ref{BiBiwf}~?
All components of the RS vector 
$\bm{F}$ satisfy the wave equation (\ref{waveq}) which can in turn be derived from the Klein-Gordon-type Lagrangian analogous to (\ref{BBKGlag}),
\begin{eqnarray}
\label{BBKGlagV}
L_F=\frac{1}{2}(\partial_\mu\bm{F}^*)\cdot(\partial^\mu \bm{F})=\frac{1}{4}\Big(
\partial_\mu\bE\cdot\partial^\mu\bE
+\partial_\mu\bB\cdot\partial^\mu\bB
\Big).
\end{eqnarray}
This Lagrangian is plainly symmetric under duality (\ref{Vdual}) with associated Noether current
\begin{eqnarray}
\label{Noethercv}
z_\mu
=\frac{1}{2}\Big((\partial_\mu \bE)\cdot\bB-(\partial_\mu\bB)\cdot\bE \Big),
\end{eqnarray} 
whose time component is a conserved charge, \begin{eqnarray}
\label{zilch}
Z=\int\! d^3\br\,\frac{1}{2}\Big((\partial_t \bE)\cdot\bB-(\partial_t\bB)\cdot\bE \Big)
=
\int\! d^3\br\,\frac{1}{2}\Big(\bB\cdot\bnabla\times\bB+\bE\cdot\bnabla\times\bE\Big),\quad
\end{eqnarray}
upon using the Maxwell equations (\ref{zilch}). This expression is reminiscent of (\ref{truehelicity})
but with field strengths instead of potentials  (consistently with (\ref{RSvector}) vs (\ref{ptF+F_})). It is in fact Lipkin's ``$Z^{000}$-\emph{zilch}" \cite{Lipkin}.
Its space part,
\begin{eqnarray}
\bm{z}=\frac{1}{2}\int d^3\br\ \Big(\bE\times(\bnabla\times\bB)-\bB\times(\bnabla\times\bE)\Big),  
\label{chiralcEBint}
\end{eqnarray}
is  in turn Lipkin's $Z^{0i0}=Z^{00i}$, identified as the \emph{optical chirality flow},  cf. eqn. \# (8.1) in \cite{BaCaYa}. 

\section{Consistency with the Weinberg-Witten theorem}

We established the duality symmetry of four different frameworks, all related to Maxwell's electromagnetism, --- and got different conserved quantities: two of them are identically zero, the two others are non-trivial. How could this come about ?
 The answer is provided by Weinberg and Witten \cite{WeWi}~:

\goodbreak
\kikezd{Theorem 1}. \textit{A theory that allows the construction of a Lorentz-covariant conserved four-current $J^\mu$ cannot contain massless particles of spin $J > 1/2$ with non-vanishing values of the conserved charge  $\displaystyle\int\! J^0 d^3\br$.}

\vskip3mm
The  \emph{Lorentz-covariant} currents (\ref{BBNc}) and (\ref{BBNc6comp}), derived from a first-order  Dirac/Weyl-type Lagrangian for spin-$1$ duly vanish.
In the quadratic Klein-Gordon-type cases the non-trivial currents (\ref{BBKGNoetherc})  and (\ref{Noethercv})  are \emph{not Lorentz covariant}, though. Consider a Lorentz boost along the $z$  axis
with parameter $v$. For the zilch
 we find, for example,
 \begin{eqnarray}
\big({z^{x}}\big)^{\prime } = z^{x}
+\gamma v\Big\{ \left( \partial _{x}E_{x}\right) E^{\prime }_{y}-\left(
\partial _{x}B_{y}\right) B^{\prime }_{x}  + \left( \partial
_{x}B_{x}\right) B^{\prime }_{y}-\left( \partial _{x}E_{y}\right) E^{\prime }_{x}\Big \}\quad
\end{eqnarray}
where  $\gamma =(\sqrt{1-v^{2}})^{-1/2}$, 
instead of $
\big({z^{x}}\big)^{\prime } = z^{x}$, as would be required for a Lorentz vector.
The ``helicity-generating current'' $j^\mu$ in (\ref{BBKGNoetherc}) behaves similarly. 
Therefore, the Weinberg-Witten theorem does \emph{not} apply to these cases  \cite{WeWi} allowing for non-zero charges --- namely optical helicity  (\ref{optihel}), or the ``zilch'', (\ref{zilch}).

\section{Conclusion}

The  concept of a ``photon wave function''  has long been debated; here we merely used it as a trick to rewrite  electromagnetism in a Dirac/Weyl resp. Klein-Gordon-type form, allowing us to use field theoretical tools.  Our trick of replacing the e.m. fields by the respective potentials  works because all components  satisfy, in the Lorentz gauge, the  wave equation (\ref{waveq}), allowing for the ``square root trick''. 
  Then the transcription  (\ref{ptF+F_}) allows us to derive the duality/helicity correspondence   mimicking the procedure used for spin $1/2$.

From our four ``variations'', the Dirac-types have vanishing helicity. 
 Our preference goes therefore to the Klein-Gordon type theory  with potentials, discussed in section \ref{potwf} and listed in the third row of  Table \ref{4theories}. It is equivalent to the double-CS-type theories advocated in \cite{BBN, BaCaYa, Camer}. When applied to the original framework of Section \ref{BiBiwf}, it yields instead Lipkin's ``zilch", which are hence also associated with duality symmetry.
 
Our findings fit perfectly into the hierarchy pattern \cite{Candlin,Kibble,BaCaYa}~: The \emph{original} theory presented in Sect. \ref{BiBiwf} is in fact obtained by replacing the potentials by their curls (i.e. the e.m. fields themselves) in the theory we propose in Sect. \ref{potwf}. The duality action (\ref{Fdual}) goes over into that on the e.m. fields, eqn. (\ref{emdual}), whereas the ``true'' helicity, (\ref{optihel}), goes over into the ``zilch'', (\ref{zilch}) \cite{Lipkin,Candlin,Kibble}.
{\small
\begin{table}
\begin{tabular}{|c|c|c|c|}
\hline
 wave function
& field equation
&Lagrangian
&conserved current and helicity
\\
\hline
\hline
$\bm{F}=\bE+i\bB$
&Dirac
$\Sigma^\mu\partial_\mu\mathcal{F}
=0,
\,
\bnabla\cdot\mathcal{F}=0$
&$\overline{\mathcal{F}}\,(\Sigma^{\mu}\partial_\mu)\,\mathcal{F}
$
&$k^\mu\equiv 0$ 
\\
\hline
$\bm{V}=\bA+i\bC$
&Dirac
$\Sigma^\mu\partial_\mu\mathcal{V}
=0,
\,
\bnabla\cdot\mathcal{V}=0$
&$\overline{\mathcal{V}}\,(\Sigma^{\mu}\partial_\mu)\,\mathcal{V}
$
&$\ell^\mu\equiv 0$ 
\\
\hline
$\bm{V}=\bA+i\bC$
&K-G
$\partial_\mu \partial^\mu \bm{V}_\pm=0,
\, \bnabla\cdot\bm{V}_\pm=0$
&$\frac{1}{2}(\partial_\mu \bm{V}_-)\cdot(\partial^\mu \bm{V}_+)$
& $j^\mu$\, helicity $\chi=\int\!\frac{1}{2}(\bA\cdot \bB - \bC\cdot\bE)
$
\\
\hline
$\bm{F}=\bE+i\bB$
&K-G
$\partial_\mu \partial^\mu \bm{F}_\pm=0,\, \bnabla\cdot\bm{F}_\pm=0$
&$\frac{1}{2}(\partial_\mu \bm{F}^*)\cdot(\partial^\mu \bm{F})$
&$z^\mu$ zilch $\int\!\!\frac{1}{2}\big(\bB\cdot\bnabla\times\bB+\bE\cdot\bnabla\times\bE\big)
$
\\
\hline
\end{tabular}\\
\caption{\it Duality-invariant wave transcriptions of electromagnetism. 
$\bm{F}$ is the Riemann-Silberstein vector and $\bm{V}$ is obtained when fields are replaced by potentials. The first-order ``Dirac-type'' transcriptions are covariant and have zero conserved current and charge. The quadratic ``Klein-Gordon-type" theories are not Lorentz-covariant and have non-trivial charges, namely helicity for $\bm{V}$, and ``zilch" for $\bm{F}$.}
\label{4theories}
\end{table}
}
 
\begin{acknowledgments}
We would like to thank J. Balog, X. Bekaert, I. Bialynicki-Birula, K. Bliokh, S. Deser, C. Duval, I.~Fernandez-Corbaton and  C. Nash for discussions and correspondence. ME and PH are grateful to the IMP of the CAS for hospitality in Lanzhou. 
This work was supported by the National Natural Science Foundation of China (Grant No. 11175215), the Major State Basic Research Development Program in China (No. 2015CB856903), and the ``Chinese Academy of Sciences President's International Fellowship Initiative (No. 2017PM0045).
\end{acknowledgments}

\goodbreak


\end{document}